\documentclass[conference]{IEEEtran}
\usepackage{cite}
\usepackage{amsmath,amssymb,amsfonts}
\usepackage{algorithmic}
\usepackage{graphicx}
\usepackage{textcomp}
\usepackage{xcolor}
\def\BibTeX{{\rm B\kern-.05em{\sc i\kern-.025em b}\kern-.08em
    T\kern-.1667em\lower.7ex\hbox{E}\kern-.125emX}}
\begin{document}

\title{Work In Progress: Towards Adaptive RF Fingerprint-based Authentication of IIoT devices.

\thanks{Identify applicable funding agency here. If none, delete this.}
}

\author{\IEEEauthorblockN{Emmanuel Lomba}
\IEEEauthorblockA{\textit{PoRTIC, Polytechnic Institute of Porto} \\
Porto, Portugal \\
emmanuel@airlomba.net}
\and
\IEEEauthorblockN{Ricardo Severino}
\IEEEauthorblockA{\textit{PoRTIC, Polytechnic Institute of Porto} \\
Porto, Portugal \\
sev@portic.ipp.pt}
\and
\IEEEauthorblockN{Ana Fernández Vilas}
\IEEEauthorblockA{\textit{Information and Computing Laboratory} \\
\textit{AtlanTTic Research Center}\\
Vigo, Spain \\
avilas@det.uvigo.es}
}

\maketitle

\begin{abstract}
As IoT technologies mature, they are increasingly finding their way into more sensitive domains, such as Medical and Industrial IoT, in which safety and cyber-security are of great importance. While the number of deployed IoT devices continues to increase exponentially, they still present severe cyber-security vulnerabilities. Effective authentication is paramount to support trustworthy IIoT communications, however, current solutions focus on upper-layer identity verification or key-based cryptography which are often inadequate to the heterogeneous IIoT environment.
In this work, we present a first step towards achieving powerful and flexible IIoT device authentication, by leveraging AI adaptive Radio Frequency Fingerprinting technique selection and tuning, at the PHY layer for highly accurate device authentication over challenging RF environments.
\end{abstract}


\section{Introduction}
The advancements in information and communication technology in the past decades have been converging into a new communication paradigm in which everything is expected to be interconnected with the heightened pervasiveness and ubiquity of the Internet of Things (IoT) archetype. As these technologies mature, they are increasingly being introduced into the industrial domain, to support what is now dubbed as the Industry 4.0, converging IoT, Cyber Physical Systems and Cloud technologies into the factory floor. However, this digitalization process is not without risks, and cyber-security is increasingly becoming the most prominent issue to be addressed in such infrastructures. 

In Industry 4.0 there may be several objectives in compromising a system, from data extraction to industrial espionage, to sabotaging and physically endanger a factory or voiding its products, or even to mount an attack to more critical parts of the same or different infrastructures~\cite{caviglione_covert_2018}. On the other hand, while the number of deployed IoT devices continues to increase exponentially and is estimated to reach 75 billion by 2025 \cite{statista_number_nodes}, IoT domains present a severe set of challenges, enhanced by the IoT scale, heterogeneity, and its fast adoption.

Indeed, IoT “things” are often equipped with limited resources in terms of memory capacity and computational power. Such limitations often hinder the direct implantation of conventional Internet security techniques like AES, or TLS into the IoT \cite{Riahi}, \cite{Kimd} and their absence may lead to various security and privacy attacks like eavesdropping, network side-channel attacks, and tracking. In addition to such heterogeneous processing capabilities, IoT encompasses a myriad of several different technologies like Wireless Sensor Networks, Radio-Frequency Identification, and Machine-to-Machine communications, which further difficult the management of dedicated cyber-security solutions.

In the IoT ecosystem, most of these Internet connected devices do not have the same experience induced resilience to intrusion, hacking and sabotage attacks that other computing devices have acquired. On the contrary, they show a signiﬁcant level of vulnerability. With 70\% of IoT devices found to have serious security vulnerabilities \cite{iotdevicesvulnerabilities}, such as unencrypted network services, weak password requirements, and 90\% of devices collecting personal information, there is a critical need for an IoT security approach capable of defeating attacks in which illegitimate devices digitally masquerade as authorized IoT devices. The need is further exacerbated as bad actors exploit this weakness to conduct attacks against other infrastructures  \cite{stanislav_hacking_2015}, \cite{simon_internet_2016}, even taking down large swaths of the Internet by leveraging D-DoS attacks such as the Mirai botnet \cite{krebs_mirai_nodate}, \cite{kolias_ddos_2017}, which relied upon illegitimate usage of hundreds of thousands of IoT devices.

All these factors impose an urgent need for an effective IoT security solution that can address a multi-dimensional problem composed of several Quality of Service (QoS) dimensions such as timeliness, scalability and heterogeneity, meeting Industry 4.0 challenges.

In conventional computing systems, the security issues of authentication, confidentiality and integrity are usually handled above the physical layer by relying on loose upper-layer identity verification (“who it is”) and key-based cryptography (“what it holds”). Regarding upper-layer identity, mechanisms usually rely in the media access control (MAC) address for identification purpose. However, this kind of authentication scheme is vulnerable to identity-based attacks, such as spoofing attack \cite{Simon} and the Sybil attack \cite{Kimd}. Moreover, as these identity-based attacks are based on changing the upper-layer identity, its contents can be easily revised, and the attack can be conveniently launched repeatedly.
Key-based cryptography is another widely used upper-layer security technique, based upon the “what it holds” strategy. Although cryptography is an effective method to defend against identity-based attacks, its application in IoT has severe limitations particularly in terms of scalability, and importantly timeliness \cite{Tiberti}, as the utilization of high complexity encryption algorithms in such wireless devices can result in large latency, which is intolerable for time-critical communications. In addition, upper-layer cryptography-based authentication is not suitable for all devices’ authentication purposes, such as for relay nodes which work at the physical layer by amplifying and forwarding a wireless signal. We must also not dismiss that vulnerabilities may be introduced into the implementation of cryptographic systems, nor that the time spent on cracking a digital security key could be shortened remarkably as available processing power increases.

Therefore, upper-layer identity verification and cryptography-based authentication schemes present drawbacks and challenges when applied to IIoT. Therefore, it is valuable to explore stable and unique physical-layer characteristics to generate device fingerprints and investigate corresponding physical-layer authentication techniques.

In our work, we intend to rely on passive authentication layers, capable of identifying the device by “who it is”, looking into characteristics that are unique in its Radio Frequency (RF) signal; \textit{i.e.}, RF Fingerprinting. Importantly, our major contribution is focused in devising and applying AI techniques capable of self tuning key aspects of the feature extraction process to greatly increase the accuracy of the authentication over different and somewhat unpredictable wireless environments. This is to be achieved via the deployment of software-defined technologies, in particular Software-Defined Radio (SDR) gateways to implement the security solutions, tackling the heterogeneity of the ecosystem, coupled with AI Edge/Cloud support, to scale and leverage Machine Learning (ML) strategies. The first step towards this goal is to develop effective and automated ways to extract the device features from its RF signal and generate the corresponding fingerprint using SDR technology. Currently in progress, in this paper we introduce our ongoing work on the automated RF feature extraction framework.





\section{Related Work}
Given the vulnerability of loose upper-layer identity verification and the limitations of key-based cryptography when deployed in IoT, particularly on timeliness, it is valuable to explore stable and unique physical-layer characteristics to generate device fingerprinting and investigate corresponding physical-layer authentication techniques.
The PHY layer approach known as Specific Emitter Identification has been put forward as a solution capable of addressing this critical IoT need \cite{Thangavelu}. One specific implementation, known as Radio Frequency Fingerprinting (RFF), facilitates radio discrimination by exploiting the unintentional ‘coloration’ that is inherently imparted upon a radio’s waveform during its generation and transmission, and has been shown feasible in multiple protocols VHF, IEEE 802.11, Bluetooth, IEEE 802.15.4 and RFID transponders.
The fact that RFF features are inherent and unique to a given radio makes them virtually impossible to imitate, thus, making security approaches based upon them difficult to bypass. Unlike bit-level security protocols, \textit{i.e.}, cryptography-based authentication techniques, RFF is proven to be a useful tool in the enhancement of wireless communications security at the physical layer in applications such as device spoofing, intrusion detection, cloning detection, indoor positioning, access control, Sybil and Replay attacks detection, to name a few. Still, the authors in \cite{Donald} show that the majority of RFF work has focused on radio classification \cite{Merchant}, where an unknown radio’s identity is determined through the comparison of its fingerprint(s) with each of the stored, reference models that represent the authorized/known radios. Such “one-to-many” comparison hinders classification, as class assignment is made no matter how poor the “best” match is. This flaw can result in the granting of network access to rogue radios. This has led to the proposal of a “one-to-one” comparison known as radio identity (ID) verification. In radio ID verification, the RF fingerprint of the unknown radio is compared only to the stored reference model associated with the presented digital ID, deeming it either authorized or rejected as a rogue.
However, proposals in RFF rely on a quite limited and inflexible subset of the signal features due to the limited available processing and computing power \cite{Donald}, which makes their accuracy highly dependent of the Signal-to-Noise Ratio (SNR). Regarding this, SDR can effectively improve the feature extraction process in complex and challenging wireless environments by careful tuning key variables such as filters bandwidth or amplifier stages gains.

Unfortunately, current approaches do not take advantage of such  SDR flexibility during feature extraction. We intend to do this by leveraging AI techniques. Also, none consider the heterogeneous IoT ecosystem and focus on a single protocol instead. Our framework will provide on-demand tools to tackle this challenge by leveraging the SDR gateway component. 

To further help in the classification process, we are to deploy an additional layer of Edge/Cloud services which will greatly improve the classification process by increasing the available computing power for multiple RF feature processing, increasing authentication accuracy. 

In the literature, authors presenting experimental results, acquire radio signals using diversified apparatus ranging from low cost SDR receivers such as the RTL-SDR dongle, up to high-end oscilloscopes with sampling rates of 20 GS/s and above or specific State-of-the-Art RF spectrum digitizers. The digitized signal samples are then either directly applied to the classifier with embedded feature extraction, or via some pre-processing stage before entering the classifier stage. In either methods, there is no evidence of any automatic control of the signal acquisition parameters. Clearly, most literature focus their research on the classification method and its performance.
To the best of our knowledge, there is no other approach like ours where the signal's features extraction uses AI to aim at the optimum signal acquisition conditions, controlling the SDR receiver parameters such as gains of amplifiers stages and filters characteristics.

As presented, most of the proposals in the literature suffer from impairments which limit the deployment, accuracy and flexibility of such solutions. In our work, we aim at investigating intelligent, practical, powerful, and flexible, “one-to-one” RF authentication techniques which: (1) rely on larger and adaptable sets of RF features, driven by the radio characteristics of the device to be identified, thus leveraging the available SDR properties. This is done by deploying AI techniques at the lower feature extraction layer to tune the SDR component and vastly improve its feature extraction quality in challenging wireless environments; (2) build upon SDR gateways to address the heterogeneity of the IoT ecosystem, enabling such authentication to be deployed into different IoT protocols; (3) increase available processing power and memory by relying on Edge/Cloud computing paradigm to offload processing intensive task from the gateways.

\section{Design and Implementation}
\begin{figure}[h!]
\centering
\includegraphics[width=8.5cm]{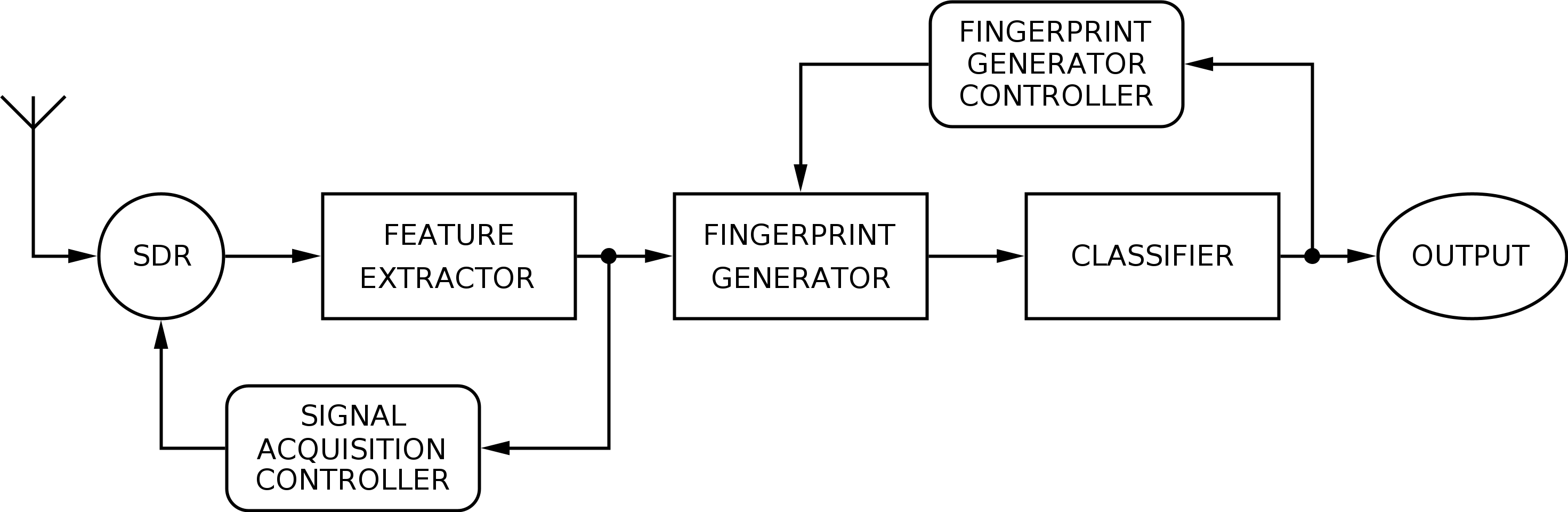}
\caption{Block diagram of the proposed architecture}
\label{im:bd}
\end{figure}

RFF-based authentication usually follows a six steps procedure: Signal acquisition, Detection of the Signal's Region of Interest, Features extraction, Features dimension reduction, Fingerprinting, and Device classification. Our RFF-based authentication approach aims at achieving powerful and flexible IoT device security, by relying upon Software-Defined Radio technology and Machine Learning (ML) at the lowest OSI-model level. 
Figure \ref{im:bd} illustrates the proposed architecture up to the classifier output. 
The feature extraction stage shall work in a closed loop where a ML algorithm at the Controller block evaluates the relevant features to be extracted, optimized by fine-tuning of the SDR parameters to address eventual signal quality degradation originated by physical constraints such as SNR variations, and other channel conditions variations. The proposed architecture also aims at addressing challenges such as fingerprint portability by simple replacement of the SDR block hardware, and scalability via ML algorithms appropriate to the study of open set of transmitting devices.

\begin{figure}[h!]
\centering
\includegraphics[width=8cm]{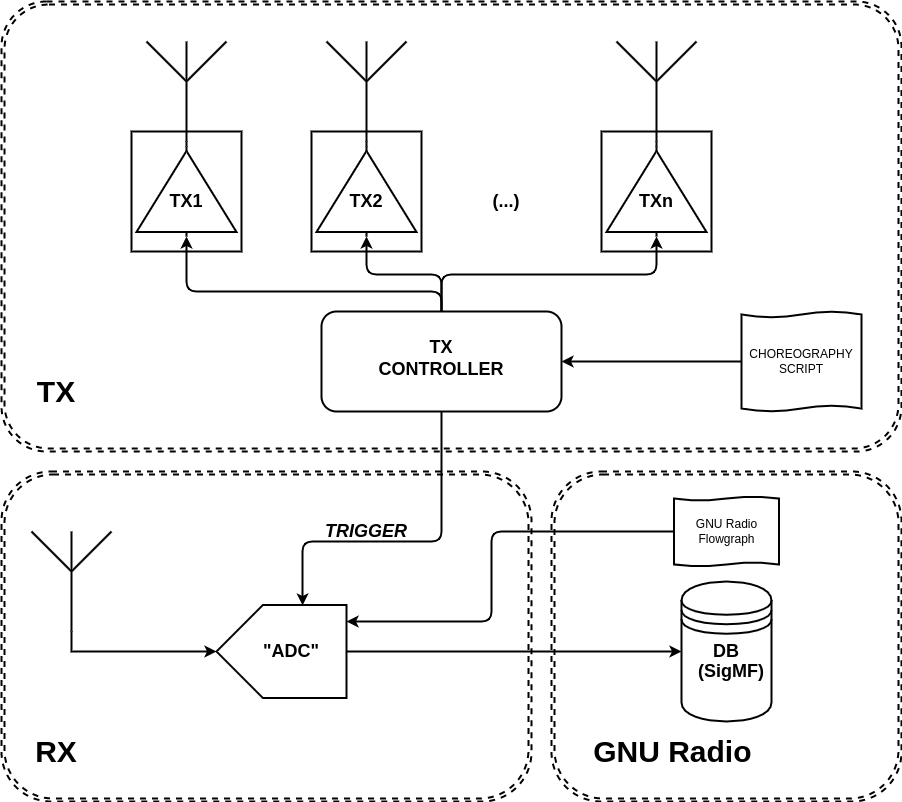}
\caption{Block diagram of the Dataset Building System}
\label{im:DBSblkdiag}
\end{figure}

To implement the first loop of the proposed architecture, an auxiliary framework is being developed, aiming at a choreographed control of several transmitters that, in sync with the SDR in the Acquisition of Signal block, will enable the production of full-featured datasets. 
Figure \ref{im:DBSblkdiag} presents the block diagram of the proposed Dataset Building System. This system is composed of two main parts: the transmitting part (TX block) and the receiving part (RX block). The transmitting block groups all the transmitters, controlled by a dedicated computer in charge of starting and ending transmissions, and triggering the acquisition of RF signals in the receiving block. Thus enabling a pseudo-continuous recording of the radio signals without risking any loss of data due to memory limitations of the SDR in the RX block; \textit{i.e.}, the same SDR to be used in the proposed architecture in Figure \ref{im:bd}.

The receiving part (RX block) is responsible for digitizing the radio signal and storing the samples into a single session file, and then saving the file into a disk for later processing (\textit{e.g.}, fingerprinting). The RX block could also be composed of a Spectrum Analyzer outputting an Intermediate Frequency to a Digitizing Oscilloscope ("ADC" in Figure \ref{im:DBSblkdiag}), just as done by the authors in \cite{6164912}, for example. In our implementation, we are comparing several low cost SDR hardware such as RTL-SDR v3.0, ADALM-Pluto and HackRF One, to be used in the RX block and later in the fingerprinting testbed. Other SDR receivers of higher performance such as USRPs are to be evaluated in the future. The GNU Radio ecosystem is the main tool used to control the SDR receiver (i.e., "ADC" in Figure \ref{im:DBSblkdiag}) and to save the collected I/Q samples into a dataset in accordance to the SigMF specification \cite{SigMF}. The resulting dataset is to be publicly released in the future.

The use of SDR technology in this work allows usability in a wide range of applications, in terms of spectrum, from ISM bands such as the 433 MHz band, up to the WiFi bands at 2.4 GHz and 5 GHz. The current multi-transmitter testbed uses 433 MHz transmitters controlled by a computer via an Arduino. The computer hosts the choreography of transmissions; \textit{e.g.}, each transmitters sends its dedicated message at pre-programmed instants in time. This approach offers endless transmissions scenarios, from single-transmitter message to multiple transmissions at once, causing interference. The position of the transmitters and the structured environment can be changed as needed to allow the study of channel impairments.

\section{Discussion}

In this paper we introduce the ongoing work regarding an RF Fingerprinting framework for the authentication of IIoT devices, which has been primarily focused on the signal acquisition and feature extraction stages.

Although RF signals are rich in features that can be exploited for the purpose of generating a device fingerprint, feature selection is a challenging problem and is dependent on the hardware capabilities of the system. Simple or primitive features can be directly measured from the acquired signal, such as the instantaneous amplitudes, peak values, or RSS. Furthermore, such measurements can be a source for derived features. The extraction of more complex features such as frequency or phase differences require dedicated processing techniques due to their stochastic nature, hence it is usual to rely on statistical methods such as Skewness, Variance or Kurtosis. While these extraction methods seem to be easier to implement, others can extract features directly from the raw signal data (\textit{i.e.}, I/Q samples), using Transformation tools such as the Hilbert-Huang Transform, the Gabor-Wigner Transform, or the Wavelet Packet Decomposition, to name only a few.

As already mentioned, the number of features is a fundamental factor in the accuracy of the classification process, hence the importance of low-latency, high availability and processing capacity computing paradigms introduced by the edge-cloud continuum. Our envisaged architectural design addresses this.

Moreover, the effectiveness of the fingerprint must comply two fundamental characteristics: unforgeable and robust; \textit{i.e.}, the fingerprints must be impossible to counterfeit and must be stable in the presence of channel variations due to device mobility or environment changes. Such variations have yet to be dealt with in a convincing way by the relevant literature. Indeed, most work done on this topic limit their classification to closed sets of transmitters \cite{9674605}. Regarding channel impairments, RFF is subject to channel impairments from diverse origins, such as signal absorption, reflection, scattering, refraction, diffraction, or signal multipath issues causing upfade, downfade, nulling, data corruption. The effects of some of these impairments are addressed by the authors in \cite{Rehman} but these have yet to be tackled in a self adaptive fashion.

Lastly, the aging of the device and the influence of temperature in the emitter operation may also interfere on some features used for fingerprinting, with a clear impact at the classification stage. All of these issues are objects of our research, which we intend to address by using AI right at the feature extraction stage, together with the SDR component, as presented in the paper.

Currently, our implementation is capable of automatically trigger the signal acquisition upon detection of a transmission and deploying a set of feature extraction modules. As we build our RF feature dataset in different wireless environments, upon different settings, we will address the development of the AI modules to carryout the adaptation of SDR elements to improve feature quality. In parallel, we are developing the Edge/cloud architecture to carryout signal classification.

\section*{Acknowledgment}
This work was partially supported by the Norte Portugal Regional Operational Programme (NORTE 2020), under the PORTUGAL 2020 Partnership Agreement, through the European Regional Development Fund (ERDF), within project "Cybers SeC IP" (NORTE-01-0145-FEDER-000044).

\end{document}